\documentclass{article}
\usepackage{amsmath}
\usepackage{amsfonts}
\usepackage{amssymb}
\usepackage{hyperref}

\usepackage{xcolor, soul}
\usepackage{soul}

\providecommand{\keywords}[1]
{
  \small	
  \textbf{\textit{Keywords---}} #1
}

\begin{document}

\title
{\textbf{Palatial Twistors from Quantum 
Inhomogeneous Conformal Symmetries and Twistorial DSR Algebras\footnote{\large Submitted to Special Issue of SYMMETRY
 ``Quantum Group Symmetry and Quanatum Geometry'', ed. A. Ballesteros, G. Gubitosi, F.J. Herranz (2021)}}}

\author{
{\textbf{Jerzy Lukierski }}
 \\
Institute for Theoretical Physics, University of Wroclaw,
\\
   pl. Maxa Borna 9, 50-205 Wroclaw, Poland
     \\
 jerzy.lukierski@uwr.edu.pl                                                                    
\\ \\
{\Large\textsl{Dedicated on 90-th Birthday to Sir Roger Penrose }}
\\   
 \Large\textsl{as a tribute to his ideas and  achievements}
 }

\date{}
\maketitle

\abstract{We construct recently introduced palatial NC twistors by considering the pair of conjugated (Born-dual) twist-deformed $D=4$ quantum inhomogeneous conformal Hopf algebras $\mathcal{U}%
_{\theta }(su(2,2)\ltimes T^{4}$) and $\mathcal{U}_{\bar{\theta}%
}(su(2,2)\ltimes \bar{T}^{4}$), where $T^{4}$ describes complex twistor coordinates
and $\bar{T}^{4}$ the conjugated dual twistor momenta. The palatial twistors are suitably chosen as the quantum-covariant modules (NC representations) of the introduced 
 Born-dual Hopf algebras. Subsequently,  we introduce the quantum deformations of $D=4$ Heisenberg-conformal algebra (HCA) $su(2,2)\ltimes H^{4,4}_\hslash$ ($H^{4,4}_\hslash=\bar{T}^4 \ltimes_\hslash T_4$ is the Heisenberg algebra of twistorial oscillators) providing in twistorial framework the basic covariant quantum elementary system. 
 The class of algebras describing deformation of HCA with dimensionfull deformation parameter, linked with Planck length $\lambda_p$, is called the twistorial DSR (TDSR) algebra, following the terminology of DSR algebra in space-time framework. 
 We   describe the examples of TDSR algebra linked with Palatial twistors which are introduced by
 the Drinfeld twist and   the quantization map in $H_\hslash^{4,4}$. We also  introduce 
 generalized quantum twistorial phase space by considering the Heisenberg double of Hopf algebra $\mathcal{U}_\theta(su(2,2)\ltimes T^4).$}


\section{Introduction}
\subsection{Towards Quantum Gravity}

One can distinguish two basic levels in quantization procedure of physical models describing contemporary fundamental
 interactions:
 
(i) The first level  can be called quantum-mechanical with canonically quantized phase space 
 coordinates and possible presence
 of classical gravity only as a static background.

On such a level,  we find all familiar relativistic quantum field theories, e.g., QED and QCD 
(fields quantized, space-time geometry flat and  Minkowskian). 

(ii) The second level also has  quantized   gravity and noncommutative space-times (all fields, including gravity
 and space-time geometry are quantized).

Quantum gravity (QG) remains   a subject of rather hypothetical models (see,   e.g., \cite{oriti,AshLew,CRove,AmJurk}), however it is mostly agreed that QG effects require at ultra-short distances the replacement of classical Einsteinian 
space-time by quantum noncommutative space-time geometry (see \cite{beggs}). The QG-generated noncommutativity corrections appear as proportional to the powers of Planck mass $m_p$ or inverse powers of Planck length $\lambda_p$.
\begin{equation}
\lambda_p=\frac{\hslash}{m_pc}=\sqrt{\frac{\hslash G}{c^3}}
\label{len}
\end{equation}
 where $c$ is the light velocity and $G$ is the gravitational Newton constant. 
 The QG origin of Planck length can be seen from   Formula (\ref{len}), with simultaneous presence of $\hslash$ and $G$. 

In order to study algebraically  the QG modifications of the space-time geometry
 in Special Relativity,  one can look at the $\lambda$-dependent deformations $\mathcal{U}_{\lambda}(\mathcal{P}^{3,1})$ of the Poincar\'{e} algebra $\mathcal{P}^{3,1}=o(3,1)\ltimes P^{3,1}$, where $P^{3,1}$ denotes the four-momenta sector
 and $\lambda$ describes an elementary length parameter which 
 can be fixed $\lambda = \lambda_p$. 
 Further, we consider the Minkowski space-time coordinates $x_\mu\in M^{3,1}$ together with covariantly acting Poincar\'{e} symmetry and introduce the semi-direct product algebra
\begin{equation}
\mathbb{A}=\mathcal{P}^{3,1}\ltimes M^{3,1}\simeq o(3,1)\ltimes(P^{3,1}\ltimes M^{3,1})
\label{222}
\end{equation}
where $P^{3,1}\ltimes M^{3,1}$ describes the relativistic phase space $\mathbb{P}^{3,1}=(M^{3,1};P^{3,1})$,  
 which after the first quantization level is 
 endowed with relativistic Heisenberg algebra structure.
 Such algebra $\mathbb{A}_{\hbar}$, also called   Heisenberg--Lorentz
 algebra, can be further deformed into quantum algebra 
 $\mathcal{U}_{\lambda}(\mathbb{A}_{\hbar})$, 
 which, to describe quantum symmetry, should have 
 Hopf algebra or Hopf-algebroid 
 structure. We stress that{{ only} $U_\lambda(\mathcal{P}^{3;1})$ can be introduced as Hopf algebra, i.e., Drinfeld quantum symmetry group \cite{17,r12}. 
 If $\hbar\neq 0$. the relativistic quantum phase space $\mathbb{P}^{3;1}_{\hbar}$ implies that $U_\lambda(\mathbb{A}_\hbar)$ 
 has the algebraic structure of Hopf algebroid \cite{alg1,alg2,jur,lsw,bp,lwss}}.
 
 Such class of deformations of algebra (\ref{222}) provides so-called DSR (Doubly Special Relativity) 
 algebra  describing quantum space-times with covariantly acting quantum symmetry. {{We} use the original name for DSR algebras
 \cite{camelia,kg,kowalski,borowpach}, however some authors use the name DSR for ``Deformed Special Relativity', which  has a vague informative content.}


 The name ``doubly'' is due to the dependence on two parameters: $c$ (light velocity) and $\lambda_p$ (Planck length, or $m_p \sim (l _p)^{-1})$.
 The first parameter, $c$, appears in the physical basis of the relativistic classical algebra $\mathbb{A}$ and 
 the second parameter, $\lambda$, determines the QG-induced modification of the algebraic structure (\ref{222}). 
 
 It was argued already in the 1930s  \cite{Bronstein}
 that QG models should at the  basic level depend on three fundamental nonvanishing constants, $c$, $\hslash$ and $G$, where $G$ can be replaced by $\lambda_p$ or $m_p$ (see (\ref{len}));
 if the cosmological constant or de Sitter radius of the Universe is finite, it introduces additional geometric parameter. {{The} variant of DSR algebra with additional de Sitter radius as additional geometric
 parameter  was called Triply Special Relativity (TSR); see \cite{KGLS}}.

 
 {The model of quantum space-time symmetries, which was an inspiration for introducing DSR algebras, is provided by the $\kappa$-deformed Poincar\'{e}--Hopf algebra \cite{luk,BalHerOL} with semi-direct product structure presented in \cite{mr} in so-called bicrossproduct basis.}

Our aim is to describe some class of quantum-deformed twistors and provide the counterpart of DSR algebra in the noncommutative framework of quantum twistors. It should be recognized here that
 there are already  several interesting papers dealing with quantum deformations of twistors and their geometries (see,   e.g., \cite{pen,kap,hann,brain,r3,lww,mp}).

\subsection{Elements of Twistor Theory}\label{sec12}

{For  more than half of a century,   Roger Penrose and his
collaborators (see,   e.g., \cite{pen,r1,r2,r4})} have promoted the idea that the space-time
manifold is a secondary geometric construction,   and   primary geometric
objects are twistors. In $D=4$ flat space-time,  twistors are introduced as
four-dimensional complex conformal spinors $t_{A}\in T^{4}$, endowed with the $D=4$
conformal-invariant pseudo-Hermitian $U(2,2)$ scalar product 
\begin{equation}
(\bar{t},t)\equiv \bar{t}_{A}\eta ^{AB}t_{B},\qquad A,B=1,2,3,4
\end{equation}%
where $\bar{t}_{A}\in \bar{T}^{4}$ are the complex-conjugated dual twistors 
and $\eta ^{AB}=(1,1,-1,-1).$ If we introduce twistors as the pair of $D=4$
Weyl spinors $(\alpha =1,2)$%
\begin{equation}
t_{A}=(\pi _{\alpha },\omega ^{\dot{\alpha}}) \label{weyl}
\end{equation}%
the alternative $U(2,2)$ frame should be used, with the metric %
\begin{equation}
\eta ^{AB}=\left( 
\begin{array}{cc}
0 & 1_2 
 \\ 
1_2 
 & 0%
\end{array}%
\right),  
\end{equation}%
leading to the formula%
\begin{equation}
(\bar{t},t)\equiv \bar{t}_{A}\eta ^{AB}t_{B}=\bar{\pi}_{\dot{\alpha}}\omega
^{\dot{\alpha}}+H.C.
\label{norm}
\end{equation}

We recall that the points $z_{\mu }$ of complex Minkowski
space-time are specified in $T^4$ by two-dimensional planes with twistor coordinates, $(\pi_\alpha, \omega^{\dot{\alpha}})$, satisfying the Cartan--Penrose incidence relation 
\begin{equation}
\omega ^{\dot{\alpha}}=iz^{\dot{\alpha}\beta }\pi _{\beta },\qquad z^{\dot{%
\alpha}\beta }=\frac{1}{2}(\sigma ^{\mu })^{\dot{\alpha}\beta }z_{\mu }.
\label{inc}
\end{equation}%
The quantum-mechanical twistors $\hat{t}_{A}$,$\hat{\bar{t}}_{A}$
on first basic quantization level 
 are provided by the oscillator-like canonical commutation relations (CCR) \cite{pen,r1}.
\begin{eqnarray}
\lbrack \hat{\bar{t}}_{A},\bigskip \hat{t}_{B}] &=&\hbar \eta _{AB}
\label{re1} \\
\lbrack \bigskip \hat{t}_{A},\bigskip \hat{t}_{B}] &=&[\hat{\bar{t}}_{A},%
\hat{\bar{t}}_{B}]=0. \label{re2}
\end{eqnarray}%
One can call $\hat{t}_A$ the twistor coordinates and $\hat{\bar{t}}_A$ the twistor momenta; they introduce the twistorial analog  of the relativistic quantum-mechanical
phase-space algebra for conformal-covariant twistorial models. We recall that one can obtain the twistor realization of 
$D=4$ conformal algebra $o(4,2)\simeq su(2,2)$ given by the bilinear products
of Quantum-Mechanical (QM) twistors $\hat{t}_{A}$,$\hat{\bar{t}}_{A}$ (see also Section \ref{sec31}).

The twistors $t_{A}\in T^{4}$ satisfying the Cartan--Penrose incidence relations (\ref{inc})
provide the geometric alternative for the description by complex Minkowski space-time geometry. The real Minkowski coordinates $x_\mu (z_\mu=x_\mu+iy_\mu)$ are obtained if the $2\times 2$ Hermitian matrix $z^{\dot{\alpha}\beta}$ is parameterized as follows
\begin{equation}
z^{\dot{\alpha}\beta}=\left(
\begin{array}{cc}
x_0+x_3 & x_1+ix_2\\
x_1-ix_2 & x_0-x_3
\end{array}
\right)
=(z^{\dot{\beta}\alpha})^*
\label{ghj}
\end{equation}
In such a case, one gets from (\ref{inc}) that $(t,t)=0$, i.e., the real Minkowski coordinates have as twistor counterparts the null twistor planes (so-called $\alpha$-planes), with vanishing norm (\ref{norm}).

For the discussion of QG effects,  it is more appropriate and realistic to consider twistors corresponding to curved space-time. The simplest examples of nonflat space-times are the ones with constant curvature $R$ or
cosmological constant $\Lambda =\pm \frac{1}{R^{2}}$, where $\Lambda >0$
for de Sitter and $\Lambda <0$ for anti-de Sitter geometries. In such a case,  the standard
quantization relations (\ref{re1}) and (\ref{re2}) for twistors should be modified, 
with respective deformation determined by the $\Lambda $-dependent antisymmetric
constant second rank twistor $I_{AB}$, called in twistor theory the infinity
 twistor (see,   e.g., \cite{r1}). In the basis (\ref{weyl}), it is given by the following formula (we choose further $\Lambda>0$)
\begin{equation}
I_{AB}=\left( 
\begin{array}{cc}
\frac{\Lambda }{6}\epsilon _{\alpha \beta } & 0 \\ 
0 & \epsilon ^{\dot{\alpha}\dot{\beta}}%
\end{array}%
\right),  \qquad I_{AB}=-I_{BA}.
\label{i11}
\end{equation}%
One can introduce in twistorial phase space $(T^{4},\bar{T}^{4})$ the
deformation of Poisson structure described by the following complex-holomorphic $%
(2,0)$ symplectic two-form \cite{r3}%
\begin{equation}
\Omega _{2}=d(I_{AB}t^{A}dt^{B})=I_{AB}dt^{A}\wedge dt^{B}=\frac{\Lambda }{6}d\pi _{\alpha
}\wedge d\pi ^{\alpha }+d\omega _{\dot{\alpha}}\wedge d\omega ^{\dot{\alpha}}
\label{two}
\end{equation}%
generated by the holomorphic $(1,0)$ Liouville one-form $\Omega
_{1}=I_{AB}t^{A}dt^{B}$, where $\Omega _{2}=d\Omega _{1}$. In dual
twistor space $\bar{T}^{4}$,  the complex-anti-holomorphic $(0,2)$ symplectic
two-form (we denote $\bar{t}_A\equiv t_A^{\dagger}=\eta_{AB}\bar{t}^B$) 
\begin{equation}
\bar{\Omega}_{2}=d(\bar{I}^{AB}\bar{t}_{A}d\bar{t}_{B})=\bar{I}%
^{AB}d\bar{t}_{A}\wedge d\bar{t}_{B}=\frac{\Lambda }{6}d\bar{\pi}^{\dot{%
\alpha}}\wedge d\bar{\pi}_{\dot{\alpha}}+d\bar{\omega}^{\alpha }\wedge d\bar{%
\omega}_{\alpha }
\label{symple}
\end{equation}
is complex-conjugated to (\ref{two}),  which  leads to the relation
\begin{equation}
\bar{I}^{AB}= \frac{1}{2} \epsilon^{ABCD} I_{CD} 
=\left( 
\begin{array}{cc}
\epsilon _{\alpha \beta } & 0 \\ 
0 & \frac{\Lambda }{6}\epsilon ^{\dot{\alpha}\dot{\beta}}%
\end{array}%
\right) .
\label{ozn}
\end{equation}
It appears that, within the framework of Hopf-algebraic quantum deformations of inhomogeneous conformal algebras $su(2,2)$, one gets separately the deformations of twistors $\hat{t}^A\in T^4$ and $\hat{\bar{t}}\in \bar{T}^4$, which   lead  to 
 the quantum-mechanical (first level) 
 quantization of symplectic structures (\ref{two}) and (\ref{symple}).
 
The deformed twistors obtained by the quantization of symplectic Poisson structures (\ref{two})--(\ref{ozn}) are  called the palatial twistors \cite{r3}. After the quantization procedure, one gets the holomorphic and anti-holomorphic noncommutativity relations
modifying (\ref{re2}) as follows ({{further}, in many formulae, we   put $\hbar=c=1$}).
\begin{equation}
\lbrack \hat{t}^{A},\hat{t}^{B}]=\hbar \bar{I}^{AB},\qquad \lbrack \hat{\bar{%
t}}_{\dot{A}},\hat{\bar{t}}_{\dot{B}}]=\hbar I_{\dot{A}\dot{B}}. \label{def}
\end{equation}

\subsection{Twist Deformations: From Space-Time \\ to Twistors}

 Let us recall 
the well-known twisting procedure (see,   e.g., \cite{r13a,r11}) of the Poincar\'{e} 
 algebra $\mathcal{U}(o(3,1)\ltimes P^{3,1})$ which is semi-dual 
 to twisted Poincar\'{e} algebra $\mathcal{U}(o(3,1)\ltimes M^{3,1})$. 
 
 For these two basic relativistic algebras (in Minkowski space-time and four-momentum space),
 one introduces the following basic Abelian twists 
\begin{eqnarray}
\bar{F} &=&\exp \left(i \frac{\lambda^2}{2}\theta ^{\mu \nu }p_{\mu
}\otimes p_{\nu }\right),  \qquad p_{\mu }\in P^{3,1} 
\label{pp1}\\
F &=&\exp \left(i \frac{1}{2\lambda ^{2}}\theta _{\mu \nu
}x^{\mu }\otimes x^{\nu }\right),  \qquad x^{\mu }\in M^{3,1}
\label{pp2}
\end{eqnarray}%
 where 
 $\theta _{\mu \nu
}=-\theta _{\nu \mu }$ are real and dimensionless; the dimensionfull deformation parameter $\lambda $ 
 representing an elementary length in QG applications may be chosen as Planck length $\lambda_p$.
If we insert the twists (\ref{pp1}) and (\ref{pp2}) into the formulae for twist-deformed Hopf algebra modules (see,   e.g., \cite{beggs,18}),  
 we get  the quantization maps
\begin{equation}
\hat{x}^{\mu }=m[\bar{F}^{-1}(\vartriangleright \otimes 1)(x^{\mu
}\otimes 1)],\qquad \hat{p}_{\mu }=m[F^{-1}(%
\vartriangleright \otimes 1)(p_{\mu }\otimes 1)],
\label{gdef}
\end{equation}%
One obtains explicitly the well-known noncommutative $\theta$-deformed coordinates called also DFR (Dopplicher, Fredenhagen, Roberts). 
 quantum space-times \cite{dfr1,dfr2}, {{where standard} parameters $\theta^{\mu\nu}_{DSR}$ in DSR
 deformation are dimensionfull---$\theta^{\mu\nu}_{DSR}= \lambda^2 \theta^{\mu\nu}$,
 with the dimensionality 
 $ [\theta^{\mu\nu}_{DSR}]= L^2$ and $ [\theta^{\mu\nu}]= L^0$.
 The choice of $\theta^{\mu\nu}$ is purely geometric, selects directions in space-time, which
 are being deformed.} We obtain from (\ref{gdef}) 
\begin{equation} 
\left[ \hat{x}^{\mu },\hat{x}^{\nu }\right] =i\lambda ^{2}\theta ^{\mu \nu }.
\label{comm11}
\end{equation}%
For the noncommutative $\theta$-deformed four-momenta we get%
\begin{equation}
\left[ \hat{p}_{\mu },\hat{p}_{\nu }\right] =\frac{i}{\lambda ^{2}}\theta
_{\mu \nu }.
\label{comm22}
\end{equation}
If we wish to obtain both deformations (\ref{comm11}) and (\ref{comm22}) inside one algebraic structure, we observe that due to Jacobi identity the canonical relations $[x_\mu,p_\nu]=i\hslash g_{\mu\nu}$ should be modified and both deformations (\ref{comm11}) and (\ref{comm22}) can be embedded together only in a Hopf algebroid (see,   e.g., \cite{alg1,alg2,jur,lsw}).

The   paper is organized as follows.
 In Section \ref{sec2}, we describe the pair of
twisted inhomogeneous $D=4$ conformal algebras $isu(2,2)\equiv su(2,2)\ltimes T^4, \bar{i}su(2,2)\equiv su(2,2)\ltimes \bar{T}^4$. Using Hopf-algebraic twisting,  we derive the NC
relations (\ref{def}) for holomorphic (chiral) and anti-holomorphic (anti-chiral) palatial twistors as the twist-generated quantum
deformations.
 If we twist the primitive
 coproducts of $su(2,2)$ generators, we can show that the twistorial NC phase
space coordinates $(\hat{T}^{4},\hat{\bar{T}}^{4})$ 
 are covariant as Hopf algebra module
under the action of respective twist-deformed inhomogeneous $su(2,2)$ algebras. 

In Section \ref{sec3}, we  
 introduce the new notion of twistorial DSR (TDSR) algebra, in particular as 
 $\Theta_{AB}$-deformed twistorial $D=4$ Heisenberg-conformal algebra 
 $su(2,2) \ltimes H^{4,4}_{\hslash}$, 
%
where $H^{4,4}_\hbar = \bar{T}^4 \ltimes_\hslash
 T^4$ 
	containing both sectors $T^4$, $\bar{T}^4$ 
 simultaneously $\Theta$-deformed. 
 For such purpose, we   
 use the quantization map 
 which follows from the quantized versions of symplectic structures (\ref{two}) and (\ref{symple}), 
 and suitably modifies the Kronecker delta in the relation (\ref{re1}).
 We   propose the twist quantization
 by a Drinfeld twist $\mathbb{F}$ 
 (see. e.g., \cite{41ad}), %
 which is 
 related with the twists (\ref{t2}) and (\ref{t1}) 
 and only in the linear approximation in $\lambda$ of the
 quantization map provides the palatial twistors.
 Further, we consider generalized twistorial quantum phase spaces defined by twisted Heisenberg doubles of $isu(2,2)$ and $\bar{i}su(2,2)$ Hopf algebras. In such a framework, following Brain and Majid, the quantum deformation of twistor geometry considered in Section \ref{sec12}
 can be obtained by cotwist quantization of inhomogeneous conformal quantum groups $ISU(2,2)$ and $\bar{I}SU(2,2)$, 
 which,  respectively,  
 are 
 Hopf-dual to twist-deformed $isu(2,2)$ and $\bar{i}su(2,2)$ quantum symmetry algebras. 

In the concluding Section \ref{sec4}, we present an outlook, with 
 directions for possible future research.

\section{Twisted Inhomogeneous \boldmath$D=4$ Conformal Algebras, Born Duality and
Palatial Twistors} \label{sec2}

\subsection{From Poincar\'{e} to Inhomogeneous Conformal Algebras} \label{sec21}
If we pass from the $D=4$ relativistic space-time description of the Universe to twistorial geometric framework, the $D=4$ Lorentz algebra $o(3,1)\simeq SL(2;\mathbb{C})\oplus \bar{SL(2;\mathbb{C}})$ 
 with the four-vectors $x_\mu,   p_\nu$ 
 is replaced by $D=4$ conformal algebra $o(4,2)\simeq su(2,2)$ with fundamental translational degrees of freedom described by twistors $(\hat{t}_A\in T^4, {\hat{\bar{t}}}_A\in \bar{T}^4)$ 
 spanning the canonical twistorial phase space. In the space-time approach,  the orbital part of Lorentz algebra generators $M_{\mu\nu}$ can be realized in terms of $D=4$ relativistic phase space variables ($x_\mu\in M^{3,1}, p_\mu \in P^{3,1})(\mu,\nu=0,1,2,3)$ as follows 
\begin{equation}
o(3,1):\qquad M_{\mu\nu}=\hat{x}_{[\mu}\hat{p}_{\nu]}.
\label{orb}
\end{equation}
{{where the} Formula (\ref{orb}) is applicable only to the spinless systems; for the extension with spin
 in the context of quantum deformations, see, e.g., \cite{lwss}}. 
 
Using the twistorial canonical oscillator algebra (\ref{re1}) and (\ref{re2}), one can analogously express in terms of twistorial phase space coordinates $(\hat{t}_A,{\hat{\bar{t}}}_A)$ the conformal 
 generators $S_{A\bar{B}}\in su(2,2) (A,{B}=1,2,3,4)$ ({{Using} the basis (\ref{weyl}) the physical 15 generators of $D=4$ conformal algebra can be expressed by the following bilinear formulas (see also \mbox{Section \ref{sec31})}). 
\begin{eqnarray*}
p_{\alpha\dot{\beta}}=\pi_\alpha\bar{\pi}_{\dot{\beta}} \qquad M_{\alpha\beta}=\pi_\alpha\omega_\beta\qquad 
M_{\dot{\alpha}\dot{\beta}}=\bar{\pi}_{\dot{\alpha}}\bar{\pi}_{\dot{\beta}}\\
D=\pi_\alpha\omega^{\dot{\alpha}}+H.C.\qquad K_{\alpha\dot{\beta}}=\omega_\alpha\bar{\omega}_{\dot{\beta}}
\end{eqnarray*}
}
\begin{equation}
S_{A{B}}=\hat{t}_A\hat{\bar{t}}_{{B}}-\frac{1}{4}(t,t)\eta_{A{B}}
\qquad \eta^{AB} S_ {AB}=0
\label{sddsa}
\end{equation}
where $S_{A{B}}$ are the $4\times 4$ pseudo-Hermitian complex matrix $u(2,2)$ generators
\begin{equation}
S_{A{B}}^\dagger=\eta_{AC}S_{C{D}}\eta_{{D}{B}}.
\end{equation}

One can introduce the pair of Poincar\'{e} groups $(\mathcal{P}_x^{3,1}=O(3,1)\ltimes M^{3,1}$, $\mathcal{P}_p^{3,1}=O(3,1)\ltimes P^{3,1})$, with $M^{3,1},P^{3,1}$ describing Minkowski coordinates and four-momenta, as two cosets related by the Fourier transform
\begin{equation}
M^{3,1}=\frac{\mathcal{P}_x^{3,1}}{o(3,1)}\ni \{x_\mu \}\qquad P^{3,1}=\frac{\mathcal{P}_p^{3,1}}{o(3,1)}\ni \{p_\mu \}.
\label{poo}
\end{equation}
The twistorial counterparts of relations (\ref{poo}) appear as follows
\begin{equation}
T^4=\frac{ISU(2,2)}{SU(2,2)}\qquad \bar{T}^4=\frac{\bar{I}SU(2,2)}{SU(2,2)}
\label{tinc}
\end{equation}
where $ISU(2,2)=SU(2,2)\ltimes T^4$ and $\bar{I}SU(2,2)= SU(2,2)\ltimes \bar{T}^4$ describe
 the semi-dual 
 pair of twistorial inhomogeneous $D=4$ conformal groups. The twistors $t_A\in T^4$ 
 describe the fundamental holomorphic conformal spinors (holomorphic twistors)    and   $\bar{t}_A\in \bar{T}^4$ the complex-conjugated anti-holomorphic ones. Alternatively, one can use the projective twistors $CP(3)$, described by the equivalence classes $t_A\sim ct_A (c\in\mathbb{C})$ in $T^4$, which parameterize the following symmetric coset of $SU(2,2)$ (see,   e.g., \cite{brain}).
\begin{equation}
CP(3)=\frac{SU(2,2)}{S(U(2,1)\otimes U(1))}.
\end{equation}
The complex Minkowski coordinates $z^{\dot{\alpha}\beta}\in CM(4)$ (see (\ref{ghj})) can be introduced as parameterizing the following complex Grassmanian
\begin{equation}
M(4)=\frac{SU(2,2)}{S(U(2)\otimes U(2))}.
\label{gras}
\end{equation}
The coset (\ref{gras}) parameterizes complex 2-planes in $T^4$ which are determined by non-parallel pairs of intersecting twistors $t_A^i (i=1,2; 
 A=1, \ldots 4; (t^1,t^2)\neq 0)$ and satisfy the pair (\ref{inc}) of Cartan--Penrose incidence relations. The complex Minkowski coordinates $z^\mu=\frac{1}{2}(\sigma^\mu)_{\dot{\alpha}\beta}z^{\dot{\alpha}\beta}$ are expressed by the pair of intersecting twistor coordinates 
 $t_A^i=(\pi^i_\alpha,{\omega}^{\dot{\alpha} i})$ as follows
\begin{equation}
z^{\dot{\alpha}\beta}=-\frac{i}{\pi^{1\alpha}\pi^2_\alpha}(\omega^{1\dot{\alpha}}\pi^{2\beta}-\omega^{2\dot{\alpha}}\pi^{1\beta}).
\end{equation}

The primary aim of the Penrose program  during the  last fifty years was to encode any curved Einsteinian space-time structure in geometric twistorial framework; in particular,  it was important to find the vocabulary permitting to translate any general relativity solution in space-time into the twistorial language. This goal was however achieved only partially, with modest hopes that the program 
 of finding the twistor formulation of general relativity theory 
 will be fully successful.
 However,  in last decade,  Roger Penrose became inspired by the idea that perhaps it is an easier
 task to construct the twistorial noncommutative version of quantum gravity. Such a view, conceptually attractive, however still faces the basic question of how   the appropriate formulation of quantum gravity in the  space-time picture would look. On the twistorial side, some first steps towards
 the construction of twistorial quantum gravity model were  provided by Penrose (see also \cite{mp}).

\subsection{Twist-Deformed Inhomogeneous Conformal
\\ 
 Hopf Algebras and Holomorphic/Anti-Holomorphic 
 \\
 Quantum Twis\-tors} \label{sec22}

Our first task  is to  show  how the relations (\ref{def}) can be obtained in the framework of
quantum deformations of inhomogeneous $D=4$
conformal algebras, with the respective 
 holomorphic twistor coordinates $t_A\in T^{4}$ or anti-holomorphic twistorial
complex momenta
 $\bar{t}_A\in \bar{T}^{4}$. 
 For such a purpose,  we consider the pair of semi-dual Hopf
algebras 
 $\mathbb{H}_0\equiv \mathcal{U}(isu(2,2))=\mathcal{U}(su(2,2)\ltimes T^{4})$
 and $\bar{\mathbb{H}}%
_0\equiv \mathcal{U}(\bar{i}su(2,2))=\mathcal{U}(su(2,2)\ltimes
 \bar{T}^{4})$,
 with Hermitian-conjugated generators in $T^4$ and $\bar{T}^4$,  but  with
 the same twistorial realization of $D=4$ conformal subalgebra $su(2,2)$. 
 Subsequently,  one gets the holomorphic and anti-holomorphic palatial twistors if the Hopf algebras 
 $\mathbb{H}_0$ and $\bar{\mathbb{H}}_0$ 
 are twisted,  respectively by the following pair of twists, 
\vspace{12pt}
\begin{eqnarray}
\mathcal{\bar{F}} &=&\exp \left(- \frac{1}{2\lambda}\bar{\Theta}^{AB}{\hat{\bar{t}}}_{A}\wedge
{\hat{\bar{t}}}_{B}\right),
\label{t2}\\
\mathcal{F} &=&\exp \left(- \frac{1}{2\lambda}\Theta_{AB}{\hat{t}}^{A}\wedge \hat{t}^{B}\right). 
\label{t1}
\end{eqnarray}%
where further, in Section \ref{sec31} we justify the same dependence of twists (\ref{t2}) and (\ref{t1}) from the elementary length parameter $\lambda$.

In the general case,  the antisymmetric numerical tensor $\Theta_{AB}$ can be chosen as complex, but,  in the case of Palatial twistors, because $\Theta_{AB}=I_{AB}$, they are real.
The pair of twists \mbox{(\ref{t2})} and (\ref{t1}) are dual under the twistorial Born map%
\begin{equation}
{\hat{t}}_{A}\longleftrightarrow {\hat{\bar{t}}}_{A},\qquad \Theta_{AB}\rightarrow \bar{\Theta}_{AB}
\label{inter}
\end{equation}%
which, as the twistorial counterpart of the Born duality map $x_\mu\leftrightarrow p_\mu$ \cite{r9,r10b}, interchanges the twistorial momenta ${\hat{\bar{t}}}_{A}$\ and the twistorial coordinates $\hat{t}_{A}$.

The conformal twists (\ref{t2}) and  (\ref{t1}) deforming, respectively, $isu(2,2)$ and $\bar{i}su(2,2)$ Hopf--Lie algebras are the twistorial counterpart of the Poincar\'{e} twists (\ref{pp1}) and (\ref{pp2})---the second one employed quite often 
 in the space-time approach
 (see \cite{r13a,r11}). Further, the pair of Formulae (\ref{gdef}) has the following counterpart in twistorial description 
\begin{equation}
\hat{t}_A^{\bar{\mathcal{F}}}=m[\bar{\mathcal{F}}^{-1}(\triangleright\otimes 1)(\hat{t}_A\otimes 1)]=(\bar{\mathcal{F}}_{(1)}^{-1}\triangleright \hat{t}_A)\bar{\mathcal{F}}^{-1}_{(2)}
\label{ct2}
\end{equation}
\begin{equation}
\hat{\bar{t}}_A^{\mathcal{F}}=m[\mathcal{F}^{-1}(\triangleright\otimes 1)(\hat{\bar{t}}_A\otimes 1)]=(\mathcal{F}_{(1)}^{-1}\triangleright \hat{\bar{t}}_A)\mathcal{F}^{-1}_{(2)}.
\label{ct1}
\end{equation}
The generators $\hat{t}_A^{\bar{\mathcal{F}}}$ describe the $\mathcal{U}_{\bar{\mathcal{F}}}(\bar{i}su(2,2))$ Hopf algebra modules and $\hat{\bar{t}}_A^{\mathcal{F}}$ span the module of Hopf algebra $\mathcal{U}_{\mathcal{F}}(isu(2,2))$.
Using the Hopf-algebraic action consistent with relation (\ref{norm})
\begin{equation}
\hat{\bar{t}}_A\triangleright \hat{t}_B=\eta_{{A}B}\qquad \hat{t}_A\triangleright \hat{\bar{t}}_B=-\eta_{A{B}}
\label{211}
\end{equation}
one gets the following explicit formulae 
 for conformal quantum twistors 
\begin{equation}
\mathcal{U}_{\bar{\mathcal{F}}}(\bar{i}su(2,2)):
 \qquad \hat{t}_A^{\bar{\mathcal{F}}}=\hat{t}_A+ \frac{1}{\lambda} \bar{\Theta}_{A}{\!}^{B}\hat{\bar{t}}_B
\label{de1}
\end{equation}
\begin{equation}
\mathcal{U}_{\mathcal{F}}(isu(2,2)):
\qquad \hat{\bar{t}}_A^{\mathcal{F}}=\hat{\bar{t}}_A-
 \frac{1}{\lambda} \Theta_{A}{\!}^{B}\hat{t}_B.
\label{de2}
\end{equation}

Using the Formula (\ref{ct2}) for $\bar{t}_A$ and (\ref{ct1}) for $t_A$, one gets additionally
\begin{equation}
\hat{t}_A^{\mathcal{F}}=\hat{t}_A\qquad \hat{\bar{t}}_A^{\bar{\mathcal{F}}}=\hat{\bar{t}}_A.
\end{equation}

We obtain the following two algebras describing twist-deformed quantum twistorial phase space coordinates $(\hat{t}_A^{\mathcal{F}},\hat{\bar{t}}_A^{\mathcal{F}})$ and $(\hat{t}_A^{\bar{\mathcal{F}}},\hat{\bar{t}}_A^{\bar{\mathcal{F}}})$ 
\begin{eqnarray}
[{\hat{\bar{t}}}_A^{\;\mathcal{{F}}},{\hat{\bar{t}}}_B^{\mathcal{\, {F}}}]=\frac{2}{\lambda}
\Theta_{AB}\qquad
 [\hat{{t}}_A^{\mathcal{{F}}},\hat{{t}}_B^{\mathcal{{F}}}]=0
 \qquad [\hat{\bar{t}}_A^{\mathcal{{F}}},\hat{t}_B^{\mathcal{{F}}}]=
 \eta_{AB}
\label{uio2}
\end{eqnarray}
\begin{eqnarray}
[\hat{t}_A^{\bar{\mathcal{F}}},\hat{t}_B^{{\bar{\mathcal{F}}}}]=\frac{2}{\lambda} \bar{\Theta}_{AB}\qquad [\hat{\bar{t}}_A^{{\bar{\mathcal{F}}}},\hat{\bar{t}}_B^{\bar{\mathcal{F}}}]=0\qquad [\hat{\bar{t}}_A^{{\bar{\mathcal{F}}}},\hat{t}_B^{{\bar{\mathcal{F}}}}]=
\eta_{AB}.
\label{uio1}
\end{eqnarray}
It should be stressed that the twistorial quantum phase space coordinates $(\hat{t}_A^{\bar{\mathcal{F}}}, \hat{\bar{t}}_A^\mathcal{F})$ satisfy the following Hermitian-conjugated algebra
\begin{equation}
\hat{t}_A^\mathcal{F}=(\hat{\bar{t}}_A^{\,\bar{\mathcal{F}}})^\dagger
\qquad \hat{\bar{t}}_A^\mathcal{F}=(\hat{{t}}_A^{\bar{\mathcal{F}}})^\dagger.
\end{equation}

The shifts described by Formulae  (\ref{de1}) and (\ref{de2}) are the examples of the Bogolyubov transformation (see \cite{bogo}) of twistorial oscillators satisfying the relations (\ref{re1}) and (\ref{re2}).

\subsection{The Twisted Conformal Covariance of 
 Quantum Twistors and Born Duality Map} \label{sec23}
The relations (\ref{uio1}) and (\ref{uio2}) are quantum-covariant  under the action of the twisted inhomogeneous conformal Hopf algebras $\mathbb{H}_\mathcal{F}=\mathcal{U}_\mathcal{F}(isu(2,2))$ and $\mathbb{H}_{\bar{\mathcal{F}}}=\mathcal{U}_{\bar{\mathcal{F}}}(\bar{i}su(2,2))$, respectively. 
In order to demonstrate such a property,  one should calculate the twist-deformed coproducts using the 
 familiar similarity maps
\begin{eqnarray}
&\Delta_{\mathcal{F}}(\hat{g})=\mathcal{F}^{-1}\circ \Delta_0(\hat{g})\circ\mathcal{F}
\qquad 
\hat{g}=(\hat{t},\hat{S}_{A\bar{B}})\in isu(2,2)
\label{ppp1}
\\
&\Delta_{{\mathcal{\bar{{F}}}}}(\hat{{g}})=\bar{\mathcal{F}}^{-1}\circ \Delta_0(\hat{{g}})\circ\bar{\mathcal{F}}
\qquad 
\hat{g}=(\hat{\bar{t}},\hat{S}_{A\bar{B}})\in\bar{i}su(2,2)
\label{ppp2}
\end{eqnarray}
where $\Delta_0(\hat{g})=\hat{g}\otimes 1+1\otimes \hat{g}$ and
 the generators 
 $\hat{S}_{A\bar{B}}\in su(2,2)$ can be represented in terms of twistor coordinates $(\hat{t}_A,\hat{\bar{t}}_A)$ (see (\ref{orb})). 
 From Formulae  (\ref{t2}), (\ref{t1}), (\ref{ppp1}) and (\ref{ppp2}),  one obtains that
\begin{equation}
\Delta_{\mathcal{F}}(\hat{t}_A)=\Delta_0(\hat{t}_A)
 \qquad 
 \Delta_{\mathcal{\bar{F}}}(\hat{\bar{t}}_A)=\Delta_0(\hat{\bar{t}}_A)
\end{equation}
and
\begin{equation}
\Delta_{\mathcal{F}}(\hat{S}_{A\bar{B}})=\Delta_0(\hat{S}_{A\bar{B}}) +
\lambda^{-1}(\Theta_B^{ D}\hat{t}_A\otimes \hat{t}_D+
\Theta^C_{ B}\hat{t}_C \otimes \hat{t}_A)
\label{ooo1}
\end{equation}
\begin{equation}
\Delta_{\bar{\mathcal{F}}}(\hat{S}_{A\bar{B}})=\Delta_0(\hat{S}_{A\bar{B}}) +
\lambda^{-1} 
(\bar{\Theta}_B^{ D}\hat{\bar{t}}_A\otimes \hat{\bar{t}}_D+\bar{\Theta}^C_{ B}\hat{\bar{t}}_C\otimes \hat{\bar{t}}_A)
\label{ooo2}
\end{equation}
Due to the modification using 
 (\ref{ooo1}) and (\ref{ooo2}) of the primitive coproducts of $S_{AB}$, 
 one can show that
\begin{equation}
\mathbb{H}_\mathcal{F}:\qquad\qquad 
 \hat{g}\triangleright([\hat{\bar{t}}_A^\mathcal{F},\hat{\bar{t}}_B^\mathcal{F}]-
 \frac{2}{\lambda}{\Theta}_{AB})=0
\label{cov11}
\end{equation}
\begin{equation}
\mathbb{H}_{\bar{\mathcal{F}}}:\qquad\qquad
\hat{\bar{g}}\triangleright([\hat{t}_A^{\bar{\mathcal{F}}},\hat{t}_B^{\bar{\mathcal{F}}}]-
 \frac{2}{\lambda}{\bar{\Theta}}_{AB})=0
\label{cov22}
\end{equation}
where we use the standard Hopf-algebraic formula defining the action $\triangleright$ of generators $\hat{h}\in\mathbb{H}$ on the products $\hat{a}\cdot\hat{b} (\hat{a},\hat{b}\in\mathbb{A})$
\begin{equation}
\hat{h}\triangleright (\hat{a}\cdot \hat{b})=(\hat{h}_{(1)}\triangleright \hat{a})(\hat{h}_{(2)}\triangleright \hat{b}).
\label{acc}
\end{equation}
where $\mathbb{A}$ is the $\mathbb{H}$-module algebra. 
We recall that,  in the case of noncommutative $\theta_{\mu\nu}$-deformed quantum space-times and quantum four-momenta (see (\ref{comm11}) and (\ref{comm22})), one gets analogously in coproducts 
$\Delta_{F}(M_{\mu\nu})$ and $\Delta_{\bar{{F}}}(M_{\mu\nu})$ the additional terms which are linear in $\theta_{\mu\nu}$ and bilinear in four-momenta (for twist (\ref{pp1})) 
or bilinear in space-time coordinates (for twist (\ref{pp2}));
 these terms are needed for the twisted quantum Poincar\'{e} invariance of the algebraic relations (\ref{comm11}) and (\ref{comm22}) (see also \cite{r13a}).
 
 For coordinate and momenta twistors,  one can consider the quantum covariance under two different inhomogeneous 
 twisted conformal
 Hopf algebras $\mathcal{U}_\mathcal{F}(isu(2,2))$ and $\mathcal{U}_{\bar{\mathcal{F}}}(\bar{i}su(2,2))$, 
 but they 
 can be mapped into each other if we supplement the twistorial Born map (\ref{inter})
 with the following exchange relation ({$\Theta_{AB}$ {in} the general case are complex, but it should be observed that $\theta_{\mu\nu}$ in both Formulae (\ref{comm11}) and (\ref{comm22}) is real and not changing under the map (\ref{inter2})}). 
\begin{equation}
\Theta_{AB} \leftrightarrow \bar{\Theta}_{AB}
\label{inter2}
\end{equation}
The superposition of maps (\ref{inter}) and (\ref{inter2}) leads to the following Born substitution rule of the twist factors (\ref{t2}) and (\ref{t1})
\begin{equation}
\mathcal{F} \leftrightarrow \bar{\mathcal{F}}
\label{inter3}
\end{equation}
The relations (\ref{inter2}) and (\ref{inter3}) describe the Born partial duality (semi-duality) of inhomogeneous conformal Hopf-algebras $\mathcal{U}_\mathcal{F}(isu(2,2))$ and $\mathcal{U}_{\bar{\mathcal{F}}}(\bar{i}su(2,2))$ (see,   e.g., \cite{msch,schoers,shoers2o}) with interchanged
 subalgebras of twistorial momenta and coordinates.

\section{Twistorial DSR Algebra as Deformed Smashed Product of \boldmath$su(2,2)$ and Twistorial Quantum Pha\-se Space} \label{sec3}

\subsection{Twistorial DSR (TDSR) Algebra} \label{sec31}

\indent 
During  2000--2001,   the notion of Double Special Relativity (DSR)  was proposed, with the postulate that the geometry of special relativity in the presence of QG corrections is modified by quantum corrections, with Planck mass 
 or Planck length 
 playing the role of mass-like deformation parameter. 
 {In fact, in Snyder model \cite{snyder}, introducing first in the literature NC quantum space-time coordinates $\hat{x}_\mu$ by means of the relation 
\begin{equation}
[\hat{x}_\mu,\hat{x}_\nu]=gM_{\mu\nu}\qquad [g]=L^2
\end{equation}
one usually assumes that $g=\beta l^2_p$ (where $\beta$ is a dimensionless constant) and one can consider Snyder model as the
 first historical example of DSR model.}
 In the Hopf-algebraic framework of quantum groups,  the general DSR algebra can be described as quantum algebra 
$\mathcal{U}_{\lambda}(\mathbb{A}_{DSR})$, where $\mathbb{A}_{DSR}$ is given by the Formula (\ref{222}).

{If we wish to introduce the twistorial counterpart of DSR theory, described by the corresponding class of twistorial DSR algebras, one should replace the algebra $\mathbb{A}_{DSR}$ (see \mbox{(\ref{222})}) by the algebra $\mathbb{A}_{TDSR}$ }
\begin{equation}
\mathbb{A}_{TDSR}=\bar{i}su(2,2)\ltimes T^4\simeq isu(2,2)\ltimes \bar{T}^4
\simeq su(2,2)\ltimes H^{4,4}_{\hbar}
\label{dsrt}
\end{equation}
where $H^{4,4}_ {\hbar}= \bar{T}^4\rtimes_\hslash T^4$ denotes the quantum twistorial phase space, described by the twistorial oscillators algebra (see (\ref{re1}) and (\ref{re2})). Subsequently, one can introduce the twistorial DSR (TDSR) algebra as described by the following quantum deformations:
\begin{equation}
\mathcal{U}_{\lambda}(\mathbb{A}_{TDSR})\equiv \mathcal{U}_{\lambda}(su(2,2)\ltimes_\hslash
H^{4,4}_\hslash
\label{tdsr}
\end{equation}

In Formula (\ref{tdsr}), we use
 the particular case of semi-direct product, called smash product (see,   e.g., \cite{klim,bcm}) of the Hopf algebra $\mathbb{H}$ and its module algebra $\mathbb{A}$
\begin{equation}
\mathcal{H}=\mathbb{H} \ltimes \mathbb{A}.
\label{hdhd}
\end{equation} 
If $h,h'\in\mathbb{H}$ and $a,b\in\mathbb{A}$, the multiplication rule in $\mathcal{H}$ is described by the following formula $(h\otimes a\in\mathcal{H};\Delta(h)\equiv h_{(1)}\otimes h_{(2)})$
\begin{equation}
(h\otimes a)\cdot (h'\otimes b)=h\cdot h'_{(1)}\otimes(a\triangleleft h'_{(2)})b
\end{equation}
which uses as input the coalgebraic sector in $\mathbb{H}$.

One can propose two ways of constructing TDSR algebra (\ref{tdsr}), in analogy with the two ways of describing the 
 relativistic quantum NC phase space in Snyder model (see \cite{gl,mm2021,m2021}):
\begin{enumerate}
\item[(1)] by proposing the quantum twistorial map 
 as given by   Formulae (\ref{de1}) and (\ref{de2}) 
 (further in this section we   link such a map with a cochain twist quantization); 
\item[(2)] by calculating for the quantum Hopf algebras $\mathcal{U}_{\lambda}(isu(2,2))$  
 and 
 \\
   $\mathcal{U}_{\lambda}(\bar{i}su(2,2))$ the Heisenberg double construction, which provides the generalized
 twistorial quantum phase space spanned by the quantum symmetry generators and the dual 
 conformal quantum matrix group coordinates ({{for} $\kappa$-deformed Poincar\'{e}-- Heisenberg double \mbox{see \cite{19,lsw};} for $\theta_{\mu\nu}$-deformed Poincar\'{e}--Heisenberg double, see  \cite{bppp,wwll}}). 
\end{enumerate}

\subsection{De Sitter Twistors and Length/Mass Dimensionalities} \label{sec32}
In order to introduce into the twistor algebra in (\ref{re1}) and (\ref{re2}) the dimensionfull parameters,  one can use the twistorial realization of conformal algebra generators $(P_{\mu},M_{(\alpha\beta)},M_{(\dot{\alpha}\dot{\beta})}, \\
K_\mu, D)$, where
\begin{equation}
[P_{\mu}]=L^{-1}\qquad[M_{(\alpha\beta)}]=[M_{(\dot{\alpha}\dot{\beta})}]=[D]=L^0,\qquad [K_\mu]=L
\label{3.4}
\end{equation}
describe the length dimensionalities $[L]=[M]^{-1}$ ($[M]$ describes the mass dimensionality). Recalling twistorial realization of conformal algebra
\begin{eqnarray}
p_{\alpha\dot{\beta}}=\pi_\alpha\bar{\pi}_{\dot{\beta}} \qquad M_{(\alpha\beta)}=(M_{(\dot{\alpha}\dot{\beta})})^\dagger=\pi_{(\alpha}\omega_{\beta)}\\
D=\pi_\alpha\omega^{\alpha}+H.C.\qquad K_{\alpha\dot{\beta}}=\omega_\alpha\bar{\omega}_{\dot{\beta}}
\end{eqnarray}
we can easily deduce that the length dimensions of Weyl spinors $(\pi_\alpha,\omega_\alpha)$,$(\bar{\pi}_{\dot{\alpha}},\bar{\omega}_{\dot{\alpha}})$ (see (\ref{weyl})) are the following
\begin{equation}
[\pi_\alpha]= [\bar{\pi}_{\dot{\alpha}}] =L^{-\frac{1}{2}}\qquad [\omega_\alpha]=[\bar{\omega}_{\dot{\alpha}}]=L^{\frac{1}{2}}
\label{dimme}
\end{equation}
In particular,  one can introduce the rescaled dimensionless $([u]=[\bar{u}]=L^0)$ twistor components as follows
\begin{equation}
u_A=\left(
\begin{array}{c}
\lambda^{\frac{1}{2}}\pi_\alpha\\
\lambda^{-\frac{1}{2}}\bar{\omega}^{\dot{\alpha}}
\end{array}
\right) \qquad
\bar{u}_A=\left(
\begin{array}{c}
\lambda^{\frac{1}{2}}\bar{\pi}_{\dot{\alpha}}\\
\lambda^{-\frac{1}{2}}\omega^\alpha
\end{array}
\right)
\end{equation}
where $\lambda$ is the fundamental length parameter. Generalizing the particular choice in 
 (\ref{i11}) and (\ref{ozn}) for palatial twistors 
 to the antisymmetric tensorial matrix $\Theta_{AB}$
\begin{equation}
\Theta_{AB}=
\left(
\begin{array}{cc}
\theta_{\alpha\gamma} & \theta_{\alpha\dot{\delta}}\\
\theta_{\dot{\beta}\gamma} & \theta_{\dot{\beta}\dot{\delta}}
\end{array}
\right)
\end{equation}
we postulate the following dimensionalities 
\begin{equation}
[\theta_{\alpha\gamma}]=L^2\qquad [\theta_{\alpha\dot{\delta}}]=[\theta_{\dot{\beta}\gamma}]=L\qquad [\theta_{\dot{\beta}\dot{\delta}}]=L^0.
\end{equation}
consistent with the assignment 
 $\Theta_{AB}= I_{AB}$.

Using (\ref{dimme}) and (\ref{t2})--(\ref{t1}),  one obtains (recall the $[\lambda]=L$) that the expression
\begin{equation}
f\equiv -2i\lambda \ln F=
\Theta_{AB} {\hat{t}}^A\wedge {\hat{t}}^B=
\nonumber
\end{equation}
\begin{equation}
\theta_{\alpha\gamma} \pi^\alpha\wedge \pi^\gamma + \theta_{\alpha\dot{\delta}}\pi^\alpha\wedge \bar{\omega}^{\dot{\delta}}+ \theta_{\dot{\beta}\gamma}\bar{\omega}^{\dot{\beta}}\wedge \pi^\gamma+\theta_{\dot{\beta}\dot{\delta}}\bar{\omega}^{\dot{\beta}}\wedge \bar{\omega}^{\dot{\delta}} 
=\lambda \Theta^{(0)}_{AB} u^A \wedge u^B
\label{factor}
\end{equation} 
where
$\theta_{\alpha \gamma}= \lambda^2 \theta^{(0)} _{\alpha \gamma}$,
$\theta_{\alpha \dot{\delta}}= \lambda \theta^{(0)} _{\alpha \dot{\delta}}$,
$\theta_{\beta \dot{\delta}}= \theta^{(0)} _{\beta \dot{\delta}}$
 and $\theta^{90)}_{AB}$ is dimensionless. It follows that 
 $[f]=L$; similarly, one gets $[\bar{f}]=L$.

Due to these numerical values of dimensionalities in front of the exponent (\ref{factor}) in Formulae  (\ref{t2}) and (\ref{t1}) 
 the numerical factor $\lambda^{-1}$ appears.  It should be observed that, contrary to the case of space-time twists (\ref{pp1}) and (\ref{pp2}), in both twistorial twists (\ref{t2}) and \mbox{(\ref{t1})}, the scaling normalization factor is the same.

\subsection{Twist Deformation of Twistors by Drinfeld Twist} \label{sec33}

One can construct the Drinfeld twist (see,   e.g., \cite{41ad,beggmaj,ebgsmaj}) 
 by multiplication of two-cocycle twists (\ref{t2}) and (\ref{t1}) in various ways, related by BCH-type formulas. Such twist can be used for twist quantization of the 
 Heisenberg-conformal algebra (\ref{dsrt}), which becomes a quasi-bialgebroid described by the smash product of the conformal $su(2,2)$ and 
 canonical twistorial Heisenberg algebra as the $su(2,2)$ module.

We   consider the following cochain twist (see,   e.g., \cite{r11}; 
 $\lambda\sim\frac{1}{m_p}$ is real).
 
\begin{equation}
\mathbb{F} =\exp \left[ 
- \frac{1}{2\lambda} (\Theta^{AB}{\hat{t}}_{A}\wedge {\hat{t}}_{B}
+\bar{\Theta}^{AB}{\hat{\bar{t}}}_{A}\wedge {\hat{\bar{t}}}_{B})\right]
\label{twiss}
\end{equation}
 In Formula (\ref{twiss}),  the exponential factor is 
 dimensionless. Because $\mathbb{F}$ does not satisfy the two-cocycle condition, the resulting twisted coproducts are not coassociative and the twist quantization will generate the quasi-bialgebroid structure. 

One gets the twist-deformed quantum twistor variables $\hat{\xi}_R=(\hat{t}_A,\hat{\bar{t}}_A) (R=1\cdots 8)$ as describing the twist quantization of Hopf algebra module 
\begin{equation}
\hat{\xi}_R^\mathbb{F}=m(\mathbb{F}^{-1}\circ(\vartriangleright
\otimes 1)\circ(\hat{\xi}_R \otimes 1)).
\label{pop}
\end{equation}
If in (\ref{pop}),  we insert 
 (\ref{211}) and (\ref{twiss}), 
 then calculate  the contribution generated by 
 the linear $\lambda$-term in $f=\ln\mathbb{F}$, we get  
\vspace{12pt}\begin{eqnarray}
\hat{t}^\mathbb{F}_A=\hat{t}_A+\lambda^{-1}\bar{\Theta}_A^{ B}\hat{\bar{t}}_B+o(\lambda^{-2})
\label{rty1}\\
\hat{\bar{t}}^\mathbb{F}_A=\hat{\bar{t}}_A-\lambda^{-1}{\Theta}_A^{ B}\hat{t}_B+o(\lambda^{-2})
\label{rty2}
\end{eqnarray}
i.e., the  linear term in the $\lambda^{-1}$ power expression gives the quantization maps analogous to 
 (\ref{de1}) and (\ref{de2}). 
 Taking only the linear term into consideration, 
we obtain the following modification of twistorial CCR (see (\ref{re1}) and (\ref{re2}))
\begin{equation}
[\hat{\bar{t}}^\mathbb{F}_A, \hat{t}^\mathbb{F}_B]=
(\eta_{AB}- 4 \lambda^{-2}\Theta_A^{ C}\bar{\Theta}_{BC})
\label{qqq1}
\end{equation}
\begin{equation}
[\hat{t}^\mathbb{F}_A,\hat{t}^\mathbb{F}_B]=2\lambda^{-1}
\bar{\Theta}_{AB}
\qquad[\hat{\bar{t}}^\mathbb{F}_A,\hat{\bar{t}}^\mathbb{F}_B]
=2\lambda^{-1}{\Theta}_{AB}.
\label{qqq2}
\end{equation}
The relations (\ref{rty1})--(\ref{qqq2}) 
 without higher order terms in $\lambda^{-1}$ can be treated as describing the quantization map for $\Theta_{AB}$-deformed twistorial Heisenberg algebra. 

The canonical conformal covariance relations for coordinate and momentum twistors $(A,B=1\dots4;\hat{S}_{A{B}}\in su(2,2))$
\begin{eqnarray}
&[\hat{S}_{A{B}},\hat{t}_C]=\eta_{{B}C}\hat{t}_A\label{320a}
\\
&[\hat{S}_{A{B}},\hat{\bar{t}}_{\bar{C}}]=-\eta_{A{C}}\hat{\bar{t}}_{\bar{B}}\label{320b}
\end{eqnarray}
 after twisting by $\mathbb{F}$ do not remain valid. 
We arrive at 
 higher order terms in {Formulae~\mbox{(\ref{rty1})} and \mbox{(\ref{rty2})}}  
 and in (\ref{320a}) and (\ref{320b}) additional terms
 which   contain, besides the $su(2,2)$ generators $S_{A{B}}$ (see (\ref{sddsa})), the bilinear products $\hat{t}_A\hat{t}_B$, $\hat{\bar{t}}_A\hat{\bar{t}}_B$, 
 which together form the set of generators of $Sp(8;R)$ algebra, which is realized linearly on the eight-dimensional real twistor space
 $\rho_R= (\hat{t}_A+\hat{\bar{t}}_A,i(\hat{t}_A-\hat{\bar{t}}_A)$. The algebra $Sp(8;R)$ is known as providing the generalization of $D=4$ conformal symmetries in the presence of tensorial central charges \cite{45b}, 
 and it  leads to 
numerous applications, e.g., in the Vasiliev higher spin algebras \cite{45c,45d}. One can conclude therefore that the twist (\ref{twiss}), 
 which simultaneously deforms both the twistorial coordinates and momenta, could be
 better adjusted to the twist quantization of inhomogeneous 
 generalized conformal algebra~$isp(8;R)=R^8\rtimes Sp(8;R)$.

We recall that, for the algebra (\ref{dsrt}) 
 and its $\mathbb{F}$-twisted quantum version,
 the coalgebraic sector can be defined only in the framework of
 quasi-bialgebroids (quasi-Hopf algebroids)~\cite{alg1,alg2,jur,lsw,bp,lwss,bupaoy}.

\subsection{Heisenberg Doubles and Generalized Twistorial Quantum Phase Space} \label{sec34}
It can be shown that, in the  $D=4$ space-time framework, both $(4+4)$-dimensional quantum phase space as well as the $(10+10)$-dimensional one which also contains 
   the Lorentz sector can be described as the Heisenberg doubles, providing various generalizations 
 and extensions of Heisenberg algebra.

 The Heisenberg double is a special example of the smash product (\ref{hdhd}), when $\mathbb{A}$ is 
 identified with the 
 dual Hopf algebra $\mathbb{H}^\star$. In such
 a case,     the 
 nondegenerate bilinear Hopf pairing $<\cdot,\cdot>:\mathbb{H}\otimes\mathbb{A}\rightarrow \mathbb{C}$ between two Hopf algebra $\mathbb{H}$ and $\mathbb{H}^\star$ is used, with the following action $\mathbb{H}\triangleright \mathbb{H}^\star$
\begin{equation}
h\triangleright a=a_{(1)}<h,a_{(2)}>.
\end{equation} 
Subsequently, one can write
\begin{equation}
h\triangleright (ab)=a_{(1)}<h_{(1)},a_{(2)}>b_{(1)}<h_{(2)},b_{(2)}>=(h_{(1)}\triangleright a)(h_{(2)}\triangleright b)
\end{equation} 
in accordance with the action (\ref{acc}) on the Hopf algebra module. One can derive in $\mathcal{H}$ the cross relations between the algebraic sectors of $\mathbb{H}$ and $\mathbb{H}^\star$ $(h\equiv h\otimes 1, a\equiv 1\otimes a)$
\begin{equation}
h\cdot a=a_{(1)}<h_{(1)},a_{(2)}>h_{(2)}
\end{equation}
 which  completes the multiplication table in $\mathbb{H}\otimes\mathbb{H}^\star$.

In applications of the  Heisenberg double $\mathbb{H} \ltimes \mathbb{H}^\star$ to
 physical models, the Hopf-algebra $\mathbb{H}$ usually describes the generalized quantum momenta,   while the   dual Hopf algebra $\mathbb{H}^\star$ provides the sector of generalized quantum positions. In the  four-dimensional space-time approach,  one obtains the generalized $(10+10)$-dimensional quantum phase space expressed as the  Heisenberg double $\mathcal{H}^{(\mathcal{P})}$ ({$\mathbb{H}=\mathcal{U}(\hat{g})$ {denotes} the enveloping Hopf--Lie algebra    and   $\mathbb{H}^\star=\mathcal{C}(\mathbb{G})$ is the Hopf algebra of functions on the Lie group manifold $\mathbb{G}$}): 
\begin{equation}
\mathcal{H}^{(\mathcal{P})}=\mathcal{U}(i\hat{o}(3,1))\ltimes\mathcal{C}(IO(3,1))
\label{tog}
\end{equation}
where $i\hat{o}(3,1)$ describes Poincar\'{e} algebra and $IO(3,1)$ the 
 dual 
 Poincar\'{e} group.
 In the quantum case, e.g., in applications to QG,
 both Hopf algebras in (\ref{tog}) can be quantum-deformed
 in a way preserving the Hopf-algebraic duality property, 
 e.g., by twisting or $\kappa$-deformation \cite{lsw,19,bppp}.

In the twistor approach, one can choose as the generalized twistorial quantum phase space the following Heisenberg double
\begin{equation}
\mathcal{H}^{(\mathcal{T})}=\mathcal{U}_{\lambda}(isu(2,2))\ltimes\mathcal{C}_{\lambda}(ISU(2,2))
\label{tog2}
\end{equation}
where the Planck length plays the role of 
 dimensionfull 
 deformation parameter.
 In particular,  one can consider in (\ref{tog2}) the twist deformations with twists $\mathcal{F},\bar{\mathcal{F}}$ (see (\ref{t2}) and \mbox{(\ref{t1})}). If we observe that the twistors 
$\hat{t}_A,\hat{\bar{t}}_A$ as well as the twists (\ref{t2}) and  (\ref{t1}) are related by the Born map (see (\ref{inter})), we obtain the following table of four Hopf algebras, describing possible twist-deformed inhomogeneous $D=4$ quantum conformal symmetries and $D=4$ inhomogeneous quantum conformal groups 
\begin{equation}
\begin{array}{ccc}
\mathcal{U}^\mathcal{F}_{\lambda}(isu(2,2)) & \overset{\hbox{Hopf}}{\underset{\hbox{duality}}\rightleftarrows} 
& \mathcal{C}^\mathcal{F}_{\lambda}(\bar{I}SU(2,2))\\
\hbox{Born}\updownarrow \hbox{duality} & & \hbox{Born}\updownarrow \hbox{duality}\\
\mathcal{U}^\mathcal{\bar{F}}_{\lambda}(\bar{i}su(2,2)) & \overset{\hbox{Hopf}}{\underset{\hbox{duality}}\rightleftarrows} & \mathcal{C}^\mathcal{\bar{F}}_{\lambda}(ISU(2,2)).
\end{array}
\end{equation}

By Hopf duality,  the twist 
 quantization of the algebra $isu(2,2)$ $(\bar{i}su(2,2))$ is mapped into the cotwist quantization of the group $ISU(2,2)$ ($\bar{I}SU(2,2)$)
 with the following properties of algebraic and coalgebraic sectors
\begin{equation}
\begin{array}{ccc}
\begin{array}{c}
\hbox{Twist quantization}\\
\hbox{of } \mathcal{U}(isu(2,2))
\end{array} & \xleftarrow\!\!\xrightarrow[\hbox{duality}]{\hbox{Hopf\,\,}} &

\begin{array}{c}
 \hbox{cotwist quantization}\\
\hbox{of } \mathbb{C}(ISU(2,2))
\end{array}
\\
\left(
\begin{array}{c}
\hbox{multiplication in}\\
\hbox{algebra not changed,}\\
\hbox{coproducts modified}
\end{array}
\right)
 & \longleftarrow\!\longrightarrow & 
\left(
\begin{array}{c}
\hbox{multiplication in}\\
\hbox{algebra modified,}\\
\hbox{coproducts unchanged}
\end{array}
\right).
\end{array}
\end{equation}

 In $D=4$,  
 the most general $(23+23)$-dimensional twistorial DSR (TDSR) algebra can be described by the deformed Heisenberg double (\ref{tog2}) 
 with (4 + 4) NC degrees of freedom in $T^4\oplus \bar{T}^4$ and (15 + 15)-dimensional conformal
 sector 
 as the subalgebra described by the $su(2,2)$ Heisenberg double.
 
 If the Hopf algebras $\mathcal{U}(isu(2,2)),\mathcal{U}(\bar{i}su(2,2))$ are twist-deformed, the cotwist-deformed algebras $\mathcal{C}(\bar{I}SU(2,2)),\mathcal{C}(ISU(2,2))$ provide noncommutative matrix entries of quantum $SU(2,2)$ group. The noncommutativity of matrix group elements $g^A_{ B}\in SU(2,2)$   is determined by RTT relations (see,   e.g., \cite{brain,r12})
\begin{equation}
R^{AC}_{ BD}g^B_{ E}g^D_{ F}=g^D_{ C}g^A_{ B}R^{BD}_{ EF}
\end{equation}
where the $R$-matrix is expressed by the following cotwist formula
\begin{equation}
R^{AC}_{ BD}=(\mathcal{F}^T\mathcal{F}^{-1})^{AC}_{ BD}.
\end{equation}
The cotwist $\mathcal{F}^{AC}_{ BD}$ dual to the twist $\mathbb{F}$ is determined by the following evaluation map:
\begin{equation}
\mathcal{F}^{AC}_{ BD}(g,g')\equiv <\mathbb{F}|g^A_{ B}\otimes g'^C_{ D}>.
\end{equation}
The noncommutative multiplication formula of cotwisted $SU(2,2)$ matrix elements is given by the formula
\begin{equation}
g\circ g'=\mathcal{F}(g_{(1)},g_{(1)})\cdot g_{(2)}\cdot g'_{(2)}\cdot\mathcal{F}^{-1}(g_{(3)},g'_{(3)}).
\label{circi}
\end{equation}
Because the twistor coordinates
 and momenta
 $T^4,\bar{T}^4$ 
 as well as the 
 complex Minkowski space-time coordinates are expressed by the element of $SU(2,2)$ group (see (\ref{tinc})--(\ref{gras})), the NC multiplication rule (\ref{circi}) defines the noncommutativity of cotwist-deformed twistor coordinates $\hat{t}_A,\hat{\bar{t}}_A$ as well as the complex 
 quantum Minkowski coordinates $\hat{z}^{\dot{\alpha}\beta}$. Further, using the cotwist-deformed multiplication rule (\ref{circi}), one can also derive the cotwist deformation of incidence relations (\ref{inc}).

\section{Outlook} \label{sec4}

The aim of this paper is the presentation of some aspects of the NC framework for quantum-deformed twistors. Our inspiration came from the paper by Penrose \cite{r3} who under the name of palatial twistors introduced the ``physical'' class of $\Theta_{AB}$-deformed dS (de Sitter) twistors, with the parameters
 $\theta_{AB}$ determined 
 geometrically by 
 real de Sitter infinity twistor $I_{AB}$ (see (\ref{i11})).

In our scheme,  we reduce the multiparameter deformations effectively to the one-Parametric ones by using the geometric degree of freedom which describes the variable Planck length or variable Planck mass. As an example,  one can provide the generalized $\kappa$-deformations depending on the constant four-vector $a_\mu$, generating the following $a_\mu$-dependent quantum space-times
\cite{kosmas}:

\begin{equation}\label{eq4.1}
[ \hat{x} _\mu,   \hat{x}_\nu ] = i(a _\mu x_\mu -a_\nu x_\mu)
\end{equation}
where $a_\mu= \lambda a^{(0)} _\mu$ ($[\lambda]= L$, $ [a^{(0)}_ \mu]= L^0$) and   the fourvector $a^{(0)}_ \mu$ 
{are chosen as} normalized, namely $(a^{(0)}_\mu)^2= -1$ for standard time-like $\kappa$-deformation,
 $a_\mu^{(0)2}=1$ for tachyonic space-like $\kappa$-deformation 
 and $a_\mu^{(0)2}=0$ 
for light cone $\kappa$-deformation.

The following are  some directions in which one can continue the studies presented in this paper:
 
\begin{enumerate} 
\item 
If we consider the twistor correspondence with complexified space-times, one should introduce the pair of dual
twistors $(t_{A},w_{A};t_{A}w^{A}=0)$\ called ambitwistors, not linked by complex conjugation (Hermitian conjugation in quantized case), which provide the description of complex null geodesies in
complexified Minkowski space $\mathbb{M}_{4}^{\mathbb{C}}$ \cite{r6,r8,r7}. In such a case, if $w^A=(\lambda^{\alpha},\mu_{\dot{\alpha}})$, one can introduce the symplectic 2-form (see (\ref{symple}))
\begin{equation}
\tilde{\Omega}_2=d(\tilde{I}^{AB}w_Adw_B)=\tilde{I}^{AB}dw_A\wedge dw_B=\frac{\lambda}{6}d\lambda_\alpha\wedge d\lambda^\alpha+d\mu^{\dot{\alpha}}\wedge d\mu_{\dot{\alpha}}
\label{syym}
\end{equation}
where $\lambda=\frac{1}{r^2}$ appears as the second cosmological constant. 
 In such a case, the curvatures $R$ and $r$ associated with twistors
 $t_{A}$ and $w_{A}$ can be different; in particular,  if $R\gg r$ ,they may provide the tool to describe two
de Sitter geometries characterizing the cosmological macroscopic distances and the ultrashort Planckian ones. It appears that the duality map $t_A\leftrightarrow w_A$, 
 which implies the interchange relation $R\leftrightarrow r$, can be linked with Born duality relation (see,   e.g., \cite{r9,r10,r10a,r10b}).
 One can speculate that the presence of the pair of dual radii (r,R) in
 ambitwistor framework can lead to the description of quantum effects
 simultaneously at ultrashort
 (radius r) and at macroscoping (radius R) cosmological distances.
 
\item {
The $D=4$ twistorial construction presented here can be quite easily generalized to $D=3$
and $D=6$ twistors, described by the $D=3$ and $D=6$ conformal groups 
$Sp(4;\mathbb{R})\simeq O(3,2)$ and $U_{\alpha }(4;\mathbb{H})\simeq O(6,2)$. 
We add that the $D=4$ conformal group $SU(2,2)$ can    also be described as the antiunitary one 
$U_{\alpha }(4; \mathbb{C})$ \cite{r13,r18}.}
In such a way,  we deal with the antiunitary family of groups
 $U_\alpha (4;\mathbb{F})$, where field F = R, C, H.
 In addition, since the 1970s,  supertwistors \cite{Fer63a} have been studied, which are a well recognized tool
 in the studies of superparticles, 
 superstrings,   supersymmetric gauge theories and supergravity.

\item {Various quantum deformations of $SU(2,2)$ and of its complexification $SL(4;\mathbb{C})$  have been used     since the 1990s  (\cite{51,52,53,54}; see also \cite{55}). One can recall that S. Zakrzewski,  
 after classifying the $D=4$ Lorentz matrices \cite{sZak}, 
   proposed the algebraic technique to classify the classical r-matrices of Poincar\'{e} algebras \cite{56} 
 After providing the classical SU(2,2) r-matrices,  it should be possible to obtain also the $r$-matrices for inhomogeneous (pseudo) unitary algebras.}

\item{Recently,    the twistorial field-theoretic approach to formulate gauge theories and gravity in twistor space  has been promoted  (see,   e.g., \cite{rose1,rose2}),
with the dynamics described by twistorial actions.
By using local twistor geometry, one obtains in a natural {way conformal gravity} \cite{rose3}; 
{the twistorial model of Einstein gravity} with non-zero cosmological constant can also be  obtained by embedding into twistorial conformal gravity 
 \cite{rose1,rose2}. 
 The formulation of QG in twistorial framework, by analogy with the approach
 presented in 
 \cite{beggs}, 
 may require  as well the  noncommutative twistorial quantum geometry.
}
\end{enumerate}

%


\sl {The paper   was  supported by Polish National Science Center, project 
 2017/27/B/ST2/01902.} 


\section*{Acknowledgements}
  {The author would like to thank Mariusz Woronowicz for the collaboration at initial stage of this paper; Andrzej Borowiec and Mariusz Woronowicz for valuable discussions; and Evgenij Ivanov and Maciek Dunajski, the organizers, respectively, of Workshop SQ'19 (Erevan, August 2019) and the Conference ``Loops meet twistors'' (Marseille-Luminy, September 2019), where some of the results presented in this article have been presented.}
\\ 



\end{document}